\documentclass[aps,prl,showpacs,twocolumn,lineno,groupedaddress]{revtex4}  
\usepackage{graphicx}  
\usepackage{dcolumn}   
\usepackage{bm}        
\usepackage{amssymb}   

\newcommand{\adzero}     {\ensuremath{\bar{ D}^0}}
\newcommand{\dst}       {\ensuremath{ D^{*}}}

\newcommand{\dstminus}  {\ensuremath{ D^{*-}}}

\newcommand{\dstst}     {\ensuremath{ D^{**}}}

\begin{document}


\title{Measurement of Semileptonic Branching Fractions of $B$ Mesons to Narrow $D^{**}$ States\\ \vspace*{0.2cm}}
%
\author{                                                                      
V.M.~Abazov,$^{35}$                                                           
B.~Abbott,$^{72}$                                                             
M.~Abolins,$^{63}$                                                            
B.S.~Acharya,$^{29}$                                                          
M.~Adams,$^{50}$                                                              
T.~Adams,$^{48}$                                                              
M.~Agelou,$^{18}$                                                             
J.-L.~Agram,$^{19}$                                                           
S.H.~Ahn,$^{31}$                                                              
M.~Ahsan,$^{57}$
% ...
\\                                                                            
\vskip 0.30cm                                                                 
\centerline{(D\O\ Collaboration)}                                             
\vskip 0.30cm                                                                 
}                                                                             
\affiliation{                                                                 
\centerline{$^{1}$Universidad de Buenos Aires, Buenos Aires, Argentina}       
\centerline{$^{2}$LAFEX, Centro Brasileiro de Pesquisas F{\'\i}sicas,         
                  Rio de Janeiro, Brazil}                                     
\centerline{$^{3}$Universidade do Estado do Rio de Janeiro,                   
                  Rio de Janeiro, Brazil}                                     
\centerline{$^{4}$Instituto de F\'{\i}sica Te\'orica, Universidade            
                  Estadual Paulista, S\~ao Paulo, Brazil}                     
\centerline{$^{5}$University of Alberta, Edmonton, Alberta, Canada,           
               Simon Fraser University, Burnaby, British Columbia, Canada,}   
\centerline{York University, Toronto, Ontario, Canada, and                    
         McGill University, Montreal, Quebec, Canada}                         
\centerline{$^{6}$Institute of High Energy Physics, Beijing,                  
                  People's Republic of China}                                 
\centerline{$^{7}$University of Science and Technology of China, Hefei,       
                  People's Republic of China}                                 
\centerline{$^{8}$Universidad de los Andes, Bogot\'{a}, Colombia}             
\centerline{$^{9}$Center for Particle Physics, Charles University,            
                  Prague, Czech Republic}                                     
\centerline{$^{10}$Czech Technical University, Prague, Czech Republic}        
\centerline{$^{11}$Center for Particle Physics, Institute of Physics,         
                   Academy of Sciences of the Czech Republic,                 
                   Prague, Czech Republic}                                    
\centerline{$^{12}$Universidad San Francisco de Quito, Quito, Ecuador}        
\centerline{$^{13}$Laboratoire de Physique Corpusculaire, IN2P3-CNRS,         
                  Universit\'e Blaise Pascal, Clermont-Ferrand, France}       
\centerline{$^{14}$Laboratoire de Physique Subatomique et de Cosmologie,      
                  IN2P3-CNRS, Universite de Grenoble 1, Grenoble, France}     
\centerline{$^{15}$CPPM, IN2P3-CNRS, Universit\'e de la M\'editerran\'ee,     
                  Marseille, France}                                          
\centerline{$^{16}$IN2P3-CNRS, Laboratoire de l'Acc\'el\'erateur              
                  Lin\'eaire, Orsay, France}                                  
\centerline{$^{17}$LPNHE, IN2P3-CNRS, Universit\'es Paris VI and VII,         
                  Paris, France}                                              
\centerline{$^{18}$DAPNIA/Service de Physique des Particules, CEA, Saclay,    
                  France}                                                     
\centerline{$^{19}$IReS, IN2P3-CNRS, Universit\'e Louis Pasteur, Strasbourg,  
                France, and Universit\'e de Haute Alsace, Mulhouse, France}   
\centerline{$^{20}$Institut de Physique Nucl\'eaire de Lyon, IN2P3-CNRS,      
                   Universit\'e Claude Bernard, Villeurbanne, France}         
\centerline{$^{21}$III. Physikalisches Institut A, RWTH Aachen,               
                   Aachen, Germany}                                           
\centerline{$^{22}$Physikalisches Institut, Universit{\"a}t Bonn,             
                  Bonn, Germany}                                              
\centerline{$^{23}$Physikalisches Institut, Universit{\"a}t Freiburg,         
                  Freiburg, Germany}                                          
\centerline{$^{24}$Institut f{\"u}r Physik, Universit{\"a}t Mainz,            
                  Mainz, Germany}                                             
\centerline{$^{25}$Ludwig-Maximilians-Universit{\"a}t M{\"u}nchen,            
                   M{\"u}nchen, Germany}                                      
\centerline{$^{26}$Fachbereich Physik, University of Wuppertal,               
                   Wuppertal, Germany}                                        
\centerline{$^{27}$Panjab University, Chandigarh, India}                      
\centerline{$^{28}$Delhi University, Delhi, India}                            
\centerline{$^{29}$Tata Institute of Fundamental Research, Mumbai, India}     
\centerline{$^{30}$University College Dublin, Dublin, Ireland}                
\centerline{$^{31}$Korea Detector Laboratory, Korea University,               
                   Seoul, Korea}                                              
\centerline{$^{32}$CINVESTAV, Mexico City, Mexico}                            
\centerline{$^{33}$FOM-Institute NIKHEF and University of                     
                  Amsterdam/NIKHEF, Amsterdam, The Netherlands}               
\centerline{$^{34}$Radboud University Nijmegen/NIKHEF, Nijmegen, The          
                  Netherlands}                                                
\centerline{$^{35}$Joint Institute for Nuclear Research, Dubna, Russia}       
\centerline{$^{36}$Institute for Theoretical and Experimental Physics,        
                  Moscow, Russia}                                             
\centerline{$^{37}$Moscow State University, Moscow, Russia}                   
\centerline{$^{38}$Institute for High Energy Physics, Protvino, Russia}       
\centerline{$^{39}$Petersburg Nuclear Physics Institute,                      
                   St. Petersburg, Russia}                                    
\centerline{$^{40}$Lund University, Lund, Sweden, Royal Institute of          
                   Technology and Stockholm University, Stockholm,            
                   Sweden, and}                                               
\centerline{Uppsala University, Uppsala, Sweden}                              
\centerline{$^{41}$Lancaster University, Lancaster, United Kingdom}           
\centerline{$^{42}$Imperial College, London, United Kingdom}                  
\centerline{$^{43}$University of Manchester, Manchester, United Kingdom}      
\centerline{$^{44}$University of Arizona, Tucson, Arizona 85721, USA}         
\centerline{$^{45}$Lawrence Berkeley National Laboratory and University of    
                  California, Berkeley, California 94720, USA}                
\centerline{$^{46}$California State University, Fresno, California 93740, USA}
\centerline{$^{47}$University of California, Riverside, California 92521, USA}
\centerline{$^{48}$Florida State University, Tallahassee, Florida 32306, USA} 
\centerline{$^{49}$Fermi National Accelerator Laboratory, Batavia,            
                   Illinois 60510, USA}                                       
\centerline{$^{50}$University of Illinois at Chicago, Chicago,                
                   Illinois 60607, USA}                                       
\centerline{$^{51}$Northern Illinois University, DeKalb, Illinois 60115, USA} 
\centerline{$^{52}$Northwestern University, Evanston, Illinois 60208, USA}    
\centerline{$^{53}$Indiana University, Bloomington, Indiana 47405, USA}       
\centerline{$^{54}$University of Notre Dame, Notre Dame, Indiana 46556, USA}  
\centerline{$^{55}$Iowa State University, Ames, Iowa 50011, USA}              
\centerline{$^{56}$University of Kansas, Lawrence, Kansas 66045, USA}         
\centerline{$^{57}$Kansas State University, Manhattan, Kansas 66506, USA}     
\centerline{$^{58}$Louisiana Tech University, Ruston, Louisiana 71272, USA}   
\centerline{$^{59}$University of Maryland, College Park, Maryland 20742, USA} 
\centerline{$^{60}$Boston University, Boston, Massachusetts 02215, USA}       
\centerline{$^{61}$Northeastern University, Boston, Massachusetts 02115, USA} 
\centerline{$^{62}$University of Michigan, Ann Arbor, Michigan 48109, USA}    
\centerline{$^{63}$Michigan State University, East Lansing, Michigan 48824,   
                   USA}                                                       
\centerline{$^{64}$University of Mississippi, University, Mississippi 38677,  
                   USA}                                                       
\centerline{$^{65}$University of Nebraska, Lincoln, Nebraska 68588, USA}      
\centerline{$^{66}$Princeton University, Princeton, New Jersey 08544, USA}    
\centerline{$^{67}$Columbia University, New York, New York 10027, USA}        
\centerline{$^{68}$University of Rochester, Rochester, New York 14627, USA}   
\centerline{$^{69}$State University of New York, Stony Brook,                 
                   New York 11794, USA}                                       
\centerline{$^{70}$Brookhaven National Laboratory, Upton, New York 11973, USA}
\centerline{$^{71}$Langston University, Langston, Oklahoma 73050, USA}        
\centerline{$^{72}$University of Oklahoma, Norman, Oklahoma 73019, USA}       
\centerline{$^{73}$Brown University, Providence, Rhode Island 02912, USA}     
\centerline{$^{74}$University of Texas, Arlington, Texas 76019, USA}          
\centerline{$^{75}$Southern Methodist University, Dallas, Texas 75275, USA}   
\centerline{$^{76}$Rice University, Houston, Texas 77005, USA}                
\centerline{$^{77}$University of Virginia, Charlottesville, Virginia 22901,   
                   USA}                                                       
\centerline{$^{78}$University of Washington, Seattle, Washington 98195, USA}  
}                                                                             
\date{July 10, 2005}

\begin{abstract}
Using the data accumulated in 2002-2004 with the D\O\ detector 
in proton-antiproton collisions at the Fermilab Tevatron collider with
centre-of-mass energy 1.96 TeV, 
the branching fractions of the decays 
$B \to \bar{D}_1^0(2420) \mu^+ \nu_\mu X$ and 
$B \to \bar{D}_2^{*0}(2460) \mu^+ \nu_\mu X$ 
and their ratio have been measured: \\	
$
{\cal B}(\bar{b} \to B) \cdot {\cal B}(B \to  \bar{D}_1^0 \mu^+ \nu_\mu X) 
\cdot {\cal B}(\bar{D}_1^0 \to \dstminus \pi^+)= 
(0.087 \pm 0.007~\mbox{\rm (stat)} \pm 0.014~\mbox{\rm (syst)})\%; \\
{\cal B}(\bar{b} \to B) \cdot {\cal B}(B \to \bar{D}_2^{*0} \mu^+ \nu_\mu X) 
\cdot {\cal B}(\bar{D}_2^{*0} \to \dstminus \pi^+) =
(0.035 \pm 0.007~\mbox{\rm (stat)} \pm 0.008~\mbox{\rm (syst)})\%;  $~and$ \\ 
( {\cal B}(B \to \bar{D}_2^{*0} \mu^+ \nu_\mu X) \cdot 
{\cal B}(\bar{D}_2^{*0} \to \dstminus \pi^+) ) / 
    ( {\cal B}(B \to  \bar{D}_1^{0} \mu^+ \nu_\mu X) \cdot 
{\cal B}(\bar{D}_1^{0} \to \dstminus \pi^+) ) = \\
0.39 \pm 0.09~\mbox{\rm (stat)} \pm 0.12~\mbox{\rm (syst)},
$ 
where the charge conjugated states are always implied.

\end{abstract}

\maketitle


This Letter describes our investigation of the properties of 
semileptonic decays of $B$ mesons to orbitally 
excited states of the $D$ meson that have small decay widths.   
In the simplest case, these states consist of a charm quark and a light quark 
in a state with orbital angular momentum equal to one.  
In the limit of a large charm quark mass
$m_c \gg \Lambda_{QCD}$, 
one doublet of states with $j=3/2~ (D_1,~D_2^{*})$ 
and another doublet with $j=1/2~ (D_0^{*}, ~D_1^{'})$ are predicted to exist, 
where the angular momentum
$j$ is the sum of the light quark spin and orbital angular momentum.
Conservation of parity and angular momentum restricts the final states 
that are allowed in the decays of these particles collectively known as \dstst ~mesons. 
The states that decay through a D-wave,
$D_1$ and $D_2^{*}$,
are expected to have small decay widths, {\cal O}(10 MeV/$c^2$), 
while the states that decay through an S-wave,
$D_0^{*}$ and $D_1^{'}$, 
are expected to be broad, {\cal O}(100 MeV/$c^2$).  

The ratio $R$ of the semileptonic branching fractions of the $B$ meson to $D_1$ and $D_2^{*}$:
\begin{eqnarray}
 R = \frac{{\cal B}(B \to D_2^{*} \ell \bar{\nu})}{{\cal B}(B \to D_1 \ell \bar{\nu})}, 
\end{eqnarray}
is one of the least model-dependent predictions of  
Heavy Quark Effective Theory (HQET) \cite{HQET} for these states. 
This ratio is expected to be equal to 1.6 in the infinite charm quark mass limit \cite{morenas},
but it can have a lower value once {\cal O}($1/m_c$) corrections 
are taken into account \cite{leibovich, ebert}. 
Together with the measurement of the corresponding ratio $R_\pi$ 
for the non-leptonic decays $B \to \dstst \pi$, 
determination of $R$ will provide important tests of HQET and 
factorization of the non-leptonic decays \cite{leibovich}.  

The narrow \dstst ~mesons have been previously studied by several experiments, most recently at Belle
\cite{belle} where the ratio $R_\pi$
was measured. 
The semileptonic decay fractions of $B$ mesons to \dstst~mesons were reported previously by 
the ARGUS \cite{argus}, CLEO \cite{cleo}, 
OPAL \cite{opal}, ALEPH \cite{aleph}, and DELPHI (as preliminary)
\cite{delphi} collaborations,
with only the latter measuring the fraction of $B \to D_2^{*} \ell \bar{\nu}$ 
and the others setting upper limits for this decay mode.


The data set used for this analysis corresponds to 
$\approx 460$ pb$^{-1}$ of integrated 
luminosity accumulated by the D\O\ detector between April 2002 and September 2004
in proton-antiproton collisions at the Fermilab Tevatron collider at
centre-of-mass energy 1.96 TeV. 
The D\O\ detector has a central tracking system consisting of a 
silicon microstrip tracker (SMT) and a central fiber tracker (CFT)~\cite{run2det}. 
Both are located within a 2~T superconducting solenoidal 
magnet and have designs optimized for tracking and 
vertexing for $|\eta|<3$  and $|\eta|<2.5$ \cite{eta}, respectively.
The SMT has a six-barrel 
longitudinal structure, each with a set of four layers arranged 
axially around the beam pipe, and interspersed with 16 radial disks. 
Silicon sensors have typical strip pitch of $50-150$ $\mu$m. 
The CFT has eight thin coaxial barrels, each supporting two doublets 
of overlapping scintillating fibers of 0.835~mm diameter.
The next layer of detection involves a preshower constructed of scintillator strips and 
a liquid-argon/uranium calorimeter. 
An outer muon system, covering $|\eta|<2$, 
consists of a layer of tracking detectors and scintillation trigger 
counters in front of 1.8~T iron toroids, followed by two similar layers 
after the toroids~\cite{run1det}. A suite of single-muon online triggers was used to record the
data set while offline only information 
from the muon and tracking systems was used in this analysis.  

Production of narrow \dstst ~mesons  in $ B \to \dstminus \pi^+ \mu^+ \nu_\mu X$ 
decay manifests itself as resonance peaks in the $\dstminus \pi^+$ 
\cite{footnote}
invariant mass spectrum. 
To perform the measurement, the semileptonic branching fractions 
of $B$ mesons to the \dstst~mesons were normalized 
to the $ B \to \dstminus \mu^+ \nu_\mu X$ process. 

Initially, a sample of $\mu^\pm \adzero$~candidates was selected by requiring a muon 
with transverse momentum $p_T^{\mu} > 2$ GeV/$c$ and $|\eta^\mu|<2$. 
\adzero~mesons were reconstructed through 
their decays into $K^+ \pi^-$.  Two tracks with 
$p_T > 0.7$ GeV/$c$ and $|\eta|<2$ were required to belong to the same jet and to 
form a common \adzero~vertex following the procedure described in detail in Ref.~\cite{dt}.
To increase the signal yield, the event selections of Ref.~\cite{dt} were relaxed 
by removing the explicit requirement that the $p_T$ of the \adzero~exceeds 5 GeV/$c$.    
In total $216870 \pm 1280~\mbox{\rm(stat)}$ $\mu^+\adzero$ candidates were found. 

\begin{figure}
\mbox{
\includegraphics[scale=0.42]{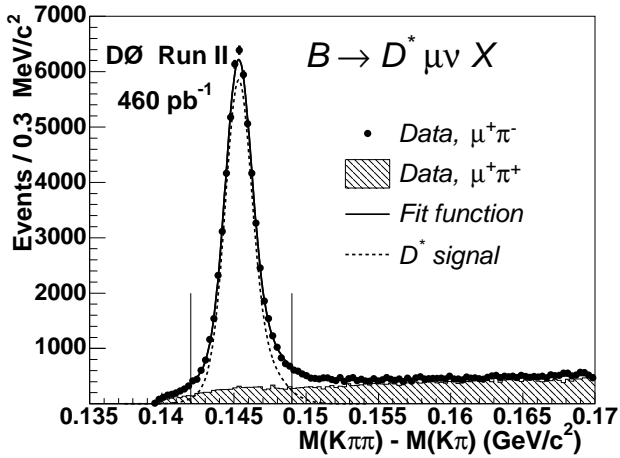}
}
\caption{\label{fig:fig2}Distribution of the mass difference $M(K\pi \pi) - M(K\pi)$
       for events with $1.75 < M(K\pi) < 1.95$ GeV/$c^2$. 
The fit function describes the signal  
as the sum of two Gaussian functions and the background 
as the sum of exponential and first-order polynomial 
functions. The signal contribution is also shown separately. 
The hatched histogram corresponds to the same-sign combination 
of the muon and pion charges.
} 
\end{figure}

\dstminus~candidates were selected through 
their decays into $\adzero \pi^-$ by requiring an additional track
with $p_T > 0.18$ GeV/$c$ and the charge opposite to that of the muon.
The mass difference $\Delta M = M(K \pi \pi) - M(K \pi)$ for all 
such tracks with assigned pion mass is shown 
in Fig.~\ref{fig:fig2} for events with $1.75 < M(K\pi) < 1.95$ GeV/$c^2$. The signal was described
as the sum of two Gaussian functions and the background 
as the sum of exponential and first-order polynomial 
functions.  
The total number of \dstminus~candidates in the peak
is $55450 \pm 280~\mbox{\rm(stat)}$ and 
is defined as the number of signal events in the mass difference window
between 0.142 and 0.149 GeV/$c^2$. 

To select a sample of $ B \to \dstminus \mu^+ \nu_\mu X$ decays used later both
for the signal search and for the normalization, 
$B$ candidates were defined using the $\mu^+$ and \dstminus~particles. 
All tracks used for the reconstruction of the $B$ candidate had to have 
at least two SMT and six CFT hits. 
The decay length of the $B$ meson, defined  
in the axial plane~\cite{axial} 
as the distance between the primary vertex~\cite{dt} and
the $B$ meson vertex, was restricted to be less than 1 cm, the uncertainty on 
the $B$-vertex axial position had to be less than 0.5 mm, and 
the $\chi^2$ of the $B$-vertex fit had to be less than 25 for three degrees of freedom. 
The significance of the decay length in the axial plane
and the proper decay length of the $B$ meson \cite{dt}
were required to exceed 3.0 and 0.25 mm respectively. The significance is defined as the ratio of 
the decay length to its uncertainty.
The last selection reduces the $c\bar{c}$ contamination in the $\mu^+ \adzero$ sample \cite{dt}. 
After these selections, the total number of \dstminus~candidates in the invariant mass difference peak
is $N_{D^*} = 31160 \pm 230~\mbox{\rm(stat)}$.

\dstst~decays can be selected by combining the \dst~candidates with an additional track with 
assigned pion mass. 
The track was required to have a charge opposite 
to that of the \dst, $p_T > 0.3$ GeV/$c$, and at least two SMT and six CFT hits. 
Pions from the \dstst~decay can also be  selected by their topology since the corresponding 
track originates from the $B$-vertex rather than from the primary vertex. 
The impact parameters (IP) in the axial plane with respect to the primary and with respect to 
the $B$-vertex were determined for each track. 
In order to select tracks belonging to the $B$-vertex, 
the ratio of IP significances of the track for the primary and 
$B$ vertices was required to be greater than four and the IP significance with 
respect to the primary vertex was required to be greater than one.
The IP significance is defined as the ratio of 
the impact parameter to its uncertainty. 

\begin{figure}
\mbox{
\includegraphics[scale=0.42]{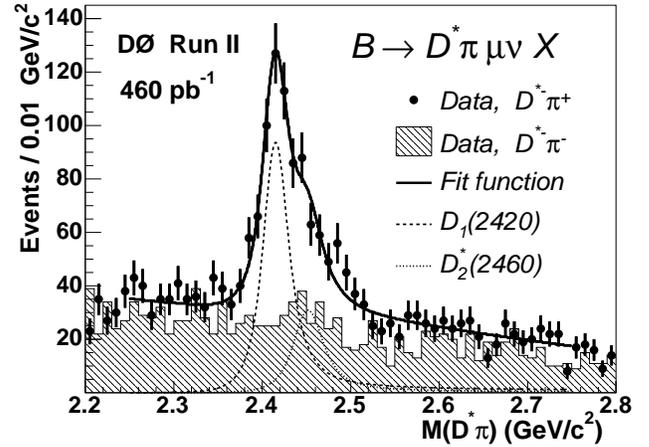}
}
    \caption[]{The invariant mass $M(\dst \pi)$.
       Event selection is described in the text. 
The points correspond to the \dst$\pi$ 
combinations with opposite charges, and the hatched histogram corresponds to the same-charge 
combinations. The distribution was fit with a sum of two 
relativistic Breit-Wigner functions corresponding 
to two narrow \dstst~states and a second-order 
polynomial describing the background.
The contributions of $D_1^0$ and $D_2^{*0}$ to the fit are shown separately.
}
    \label{fig:fig3}  
\end{figure}

The \dstst~invariant mass distribution after all selections is 
shown in Fig.~\ref{fig:fig3} where 
the \dst~mass from the Particle Data Group (PDG) \cite{pdg} 
has been used as a mass constraint. 
The observed mass peak can be interpreted as two merged narrow \dstst~states, 
$~D_1^0$ and $~D_2^{*0}$. 

The distribution was fit using a sum of two 
relativistic Breit-Wigner functions $G_{D_1}$ and $G_{D^{*}_2}$, 
corresponding to the two narrow \dstst~states with the numbers of 
events $N_{D_1}$ and $N_{D^{*}_2}$
respectively, and a second-order 
polynomial $f_{b}$ describing the background, see Eq.~(2). 

\begin{eqnarray}
f(x) & = & N_{D_1} \cdot G_{D_1} + N_{D^{*}_2} \cdot G_{D^{*}_2} +  f_{b}(x);  \\
G_i &  = & 
\frac{M_i~ \Gamma(x)}{({x}^{2}-M_i^2)^2 + M_i^2 \Gamma(x)^2} \otimes {\rm Res}_i; \nonumber \\
\Gamma(x)      &  = & \Gamma_i
 \cdot \frac{M_i}{x} \left(\frac{k}{k_0}\right)^{2L+1} \cdot F^{(L)}(k,k_0); \nonumber \\
F^{(2)}(k,k_0) &  = & \frac{9+3(k_0 z)^2+(k_0 z)^4}{9+3(k z)^2+(k z)^4}.  \nonumber 
\end{eqnarray}

In the formulas above, $x$ is the \dst$\pi$ invariant mass and 
$M_i$ and $\Gamma_i$ are the mass and width of the corresponding resonance. 
The variables $k$ and $k_0$ are 
the pion three-momenta in the \dstst~rest frame when the \dstst~has
a four-momentum-squared equal to $x^2$ and $M_i^2$, respectively.
$F^{(2)}(k,k_0)$ is the Blatt-Weisskopf form factor for D-wave ($L=2$) decays of \dstst~mesons 
\cite{blwe}, and $z = 1.6$ (GeV/$c)^{-1}$ is a hadron scale corresponding to the case of the charm quark. 
${\rm Res}_i$ is the mass resolution function described by two Gaussian functions 
with the parameterization determined 
from Monte Carlo (MC) simulations. The second Gaussian describing the resolution 
corresponded to 28\% of events and was wider by a factor of 2.2 than the first one. 
The standard D\O\ simulation chain included the
{\sc evtgen} \cite{evtgen} generator interfaced to {\sc pythia} \cite{PYTHIA} and followed 
by full {\sc geant} \cite{geant} modeling of the detector response and event reconstruction. 
The MC resolution was scaled up by 20\% to account for the 
the difference between the data and the MC, where 
the scaling factor was estimated by comparing the \dst~mass resolution in the data and MC.     
The mass resolution used for the fit (sigma of the first Gaussian in the resolution function) 
was 8.2 MeV/$c^2$ for $D^{0}_1$ and 9.4 MeV/$c^2$ for $D^{*0}_2$ after the scaling.

The parameters of the background function were determined by 
fitting the distribution of the same-charge combinations, fixing the function shape parameters,
and allowing the overall normalization 
of the background to float in the mass fit. 
The mass difference between $D^{0}_1$ and $D^{*0}_2$ and the widths of $D^{0}_1$ and 
$D^{*0}_2$ were fixed in the fit using their corresponding values from the 
PDG \cite{pdg}: 
$36.7 \pm 2.7$ MeV/$c^2$, $18.9^{+4.6}_{-3.5}$ MeV/$c^2$, $23 \pm 5$ MeV/$c^2$. 
The numbers of events in the two narrow states derived from the fit,  
$N_{D_1} = 467 \pm 39~\mbox{\rm(stat)}$ and $N_{D^{*}_2} = 176 \pm 37~\mbox{\rm(stat)}$,  
were used to determine the total number of events  
$N_{\dstst} = 643 \pm 38~\mbox{\rm(stat)}$, and the ratio 
$N_{D^{*}_2} / N_{D_1} = 0.378 \pm 0.086~\mbox{\rm(stat)}$, 
where the uncertainties take into account 
correlation between the variables. 
The $\chi^2$ of the fit at the minimum is 46.9 for 46 degrees of freedom.

The branching fraction for the decays $B \to \bar{D}^{**} \mu^+ \nu_\mu  X$ 
can be determined by normalization to the known value of 
the branching fraction 
${\cal B}(\bar{b} \to   \dstminus \ell^+ \nu X) = (2.75 \pm 0.19)\%$ \cite{pdg}. 
The following two formulas were used for the calculations:

\begin{eqnarray}
{\cal B}(\bar{b} \to B) \cdot {\cal B}(B \to \bar{D}^{**} \mu^+ \nu_\mu  X) 
\cdot {\cal B}(\bar{D}^{**} \to \dstminus \pi^+) = \nonumber \\
{\cal B}(\bar{b} \to   \dstminus \ell^+ \nu X) 
\cdot \frac{N_{\dstst}}{N_{D^*}} \cdot \frac{1}{\epsilon_{\dstst}}; \nonumber \\
\frac{{\cal B}(B \to  \bar{D}_2^{*0} \mu^+ \nu_\mu X) \cdot 
{\cal B}(\bar{D}_2^{*0} \to \dstminus \pi^+)}
     {{\cal B}(B \to  \bar{D}_1^{0} \mu^+ \nu_\mu X) \cdot 
{\cal B}(\bar{D}_1^{0} \to \dstminus \pi^+)} = 
\frac{N_{D_2^{*}}}{N_{D_1}} \cdot \frac{\epsilon_{D_1}}{\epsilon_{D_2^{*}}} \nonumber. 
\end{eqnarray}

$N_{D^{**}}$ and $N_{D^*}$
are the numbers of \dstst~and \dst~candidates
as defined above. The \dstst~notation stands for $D_1^{0}$ or $D_2^{*0}$ 
or both of them, $(D_1^{0},D_2^{*0})$. 
$\epsilon_{D^{**}}$ is the efficiency 
to reconstruct the charged pion from the \dstst~decay
determined from the MC and equal to $(47.2 \pm 1.0~\mbox{\rm(stat)})\%$ for
the $D_1^0$ meson and $(45.4 \pm 1.2~\mbox{\rm(stat)})\%$ for the $D_2^{*0}$ meson. 
Contributions from $B_s$ mesons, $\Lambda_b$ baryons, $B \to D^{(*)}_{(s)}D^{*-}X$ decays 
and the $c\bar{c}$ process to the sample were found to be small 
and have been neglected \cite{dt}.

The relative systematic uncertainties on the branching fractions 
and of their ratio are summarized in Table \ref{tab:tab2}.
\begin{table}
\caption{\label{tab:tab2} Relative systematic uncertainties on the semileptonic branching fraction to 
both narrow states, ${\cal B}_{D^{**}}$; the semileptonic branching 
fractions to $D^0_1$ and to $D^{*0}_2$, ${\cal B}_{D_1}$ and 
${\cal B}_{D^{*}_2}$; and their ratio ${\cal B}_{D^{*}_2}/{\cal B}_{D_1}$. 
}
\begin{ruledtabular}
\begin{tabular}{lcccc}
     Source                  & ${\cal B}_{D^{**}}$ & ${\cal B}_{D_1}$ &  ${\cal B}_{D^{*}_2}$ & ${\cal B}_{D^{*}_2}/{\cal B}_{D_1}$    \\ 
\hline
${\cal B}(\bar{b} \to \dstminus \ell^+ \nu X)$        & 7\%        & 7\%            & 7\%        & --    \\ 
Mass resolution              & 2\%        & 4\%            & 5\%       & 8\%    \\
$\Gamma_{D_1}$               & 3\%        & 11\%     & 16\%   & 24\%    \\ 
$\Gamma_{D_2^*}$& 2\%        & 2\%            & 11\%        & 13\%    \\ 
$\Delta M$      & 1\%        & 3\%            & 6\%  & 9\%    \\ 
MC statistics                & 2\%        & 2\%            & 2\%        & 3\%    \\
$\epsilon_{D^{**}}$ modeling         & 3\%        & 3\%            & 3\%        & 2\%    \\
Wide resonance               & 5\%        & 5\%            & 5\%        & 2\%    \\
Interference effects         & 2\%        & 2\%            & 2\%        & 2\%    \\
 \dst~fit and \dstst~bkg fit & 4\%        & 4\%            & 4\%        & 1\%    \\
\hline
Total uncertainty  &  12\%  & 16\%           &  24\% &  30\%   \\ 
\end{tabular}
\end{ruledtabular}
\end{table}
The contribution due to uncertainty in ${\cal B}(\bar{b} \to \dstminus \ell^+ \nu X)$ 
was determined from the uncertainty on this branching fraction. 
The systematic uncertainty caused by the MC mass 
resolution was estimated by varying the resolution by $\pm$20\%.
Contributions due to limited knowledge of the \dstst~masses and 
widths were computed by refitting the mass distribution after
varying these parameters within their uncertainties. 
The systematic uncertainty due to efficiency modeling accounts for the variation 
caused by a possible mismatch between the $p_T$ spectra of reconstructed particles in the data and MC.

There are predictions and possibly observations \cite{belle} 
of a wide resonance $\bar{D}_1^{'0}$ with a mass of 2430 MeV/$c^2$ and a width of 380 MeV/$c^2$ predominantly 
decaying to $D^{*-}\pi^+$. 
This resonance is not apparent in 
our data, and it was not used in the fits. 
The systematic uncertainty caused by a possible contribution of 
this resonance was evaluated 
allowing for another Breit-Wigner function in the fit, with 
the mass and width fixed to the wide resonance parameters. 

Any interference effects between the $D_1^0$ and $D_2^{*0}$ must average to zero after integration 
over all angles 
under the assumption of equal acceptances. The validity of this assumption has been checked and 
the corresponding uncertainty assigned.
The systematic uncertainty due to the fitting procedure was estimated by varying the functions describing
the 
backgrounds for the \dst~and \dstst~mass distributions 
and also the function describing the \dst~mass peak. 
The total systematic uncertainty was found by summing all the above sources in quadrature.

Using the numbers defined above, the semileptonic 
branching fractions of $B$ mesons to \dstst~mesons and their ratio are:

\begin{widetext}
\begin{eqnarray}
{\cal B}(\bar{b} \to B) \cdot {\cal B}(B \to  (\bar{D}_1^0, \bar{D}_2^{*0}) \mu^+ \nu_\mu X) 
\cdot {\cal B}((\bar{D}_1^0, \bar{D}_2^{*0}) \to \dstminus \pi^+) &=& 
(0.122 \pm 0.007~\mbox{\rm(stat)} \pm 0.015~\mbox{\rm(syst)})\%; \nonumber \\
{\cal B}(\bar{b} \to B) \cdot {\cal B}(B \to  \bar{D}_1^0 \mu^+ \nu_\mu X) 
\cdot {\cal B}(\bar{D}_1^0 \to \dstminus \pi^+)&=& 
(0.087 \pm 0.007~\mbox{\rm(stat)} \pm 0.014~\mbox{\rm(syst)})\%; \nonumber \\
{\cal B}(\bar{b} \to B) \cdot {\cal B}(B \to \bar{D}_2^{*0} \mu^+ \nu_\mu X) 
\cdot {\cal B}(\bar{D}_2^{*0} \to \dstminus \pi^+)&=& 
(0.035 \pm 0.007~\mbox{\rm(stat)} \pm 0.008~\mbox{\rm(syst)})\%; \nonumber \\ 
\frac{{\cal B}(B \to \bar{D}_2^{*0} \mu^+ \nu_\mu X) \cdot 
{\cal B}(\bar{D}_2^{*0} \to \dstminus \pi^+)}
     {{\cal B}(B \to  \bar{D}_1^{0} \mu^+ \nu_\mu X) \cdot 
{\cal B}(\bar{D}_1^{0} \to \dstminus \pi^+)} &=& 
0.39 \pm 0.09~\mbox{\rm(stat)} \pm 0.12~\mbox{\rm(syst)}. \nonumber
\end{eqnarray}
\end{widetext}

Upon using the input ${\cal B}(\bar{b} \to B) = (39.7 \pm 1.0)\%$ \cite{pdg}, assuming 
isospin conservation and that 
the $D_1$ meson decays only into $\dst\pi$ \cite{manohar}, the branching fraction 
${\cal B}(B \to  \bar{D}_1^{0} \ell^+ \nu X) = (0.33 \pm 0.06)\%$ is determined. 
It is different from the PDG value, $(0.74 \pm 0.16)\%$ \cite{pdg}, by 2.5 standard deviations. 
Similarly the branching fraction 
${\cal B}(B \to  \bar{D}_2^{*0} \ell^+ \nu X) = (0.44 \pm 0.16)\%$ is determined assuming
that $D_2^{*}$ decays into $\dst\pi$ in $(30 \pm 6)\%$ of the cases \cite{pdg}.
This result is in agreement with the 95\% CL upper limit 0.65\% provided by the PDG. 

Using the measured ratio of the branching fractions and the same assumptions 
for the absolute fractions for $D_1$ and $D_2^{*}$ as above, the ratio 
$R  =  1.31 \pm 0.29~\mbox{\rm(stat)} \pm 0.47~\mbox{\rm(syst)}$ 
was computed. 

In summary, using 460 pb$^{-1}$ of integrated luminosity accumulated with the D\O\ detector, 
the semileptonic decays  
$B \to   \bar{D}_1^0 \mu^+ \nu_\mu X$ and $B \to  \bar{D}_2^{*0}\mu^+ \nu_\mu  X$ 
have been observed and the branching fractions measured using statistics 
more than an order of magnitute better
than previous measurements \cite{argus, cleo, opal, aleph, delphi}. 
This result represents a significant improvement in the knowledge of 
$B$ branching fractions to orbitally excited $D$ mesons, 
and the first direct measurement of their ratio.

%
We thank the staffs at Fermilab and collaborating institutions, 
and acknowledge support from the 
DOE and NSF (USA);
CEA and CNRS/IN2P3 (France);
FASI, Rosatom and RFBR (Russia);
CAPES, CNPq, FAPERJ, FAPESP and FUNDUNESP (Brazil);
DAE and DST (India);
Colciencias (Colombia);
CONACyT (Mexico);
KRF (Korea);
CONICET and UBACyT (Argentina);
FOM (The Netherlands);
PPARC (United Kingdom);
MSMT (Czech Republic);
CRC Program, CFI, NSERC and WestGrid Project (Canada);
BMBF and DFG (Germany);
SFI (Ireland);
Research Corporation,
Alexander von Humboldt Foundation,
and the Marie Curie Program.
%


\begin{thebibliography}{99}
%
\bibitem[*]{lehner}
Visitor from University of Zurich, Zurich, Switzerland.
%
\vskip 0.25cm

\bibitem{HQET}
N.~Isgur and M.~B.~Wise, Phys.\ Lett. {\bf B~232}, 113 (1989); 
E.~Eichten and B.~Hill, Phys.\ Lett. {\bf B~234}, 511 (1990);
H.~Georgi, Phys.\ Lett. {\bf B~240}, 447 (1990). 
\bibitem{morenas}
V.~Morenas, A.~Le Yaouanc, L.~Oliver, O.~Pene, and J.-C.~Raynal,
Phys. Rev. D {\bf 56}, 5668 (1997);
Ming-Qiu Huang and Yuan-Ben Dai, Phys. Rev. D {\bf 59}, 034018-1 (1999). 
\bibitem{leibovich}
A.~K.~Leibovich, Z.~Ligeti, I.~W.~Stewart, and M.~B.~Wise, 
Phys. Rev. Lett. {\bf 78}, 3995 (1997); 
Phys. Rev. D {\bf 57}, 308 (1998).
\bibitem{ebert}
D.~Ebert, R.~N.~Faustov and V.~O.~Galkin, Phys. Rev. D {\bf 62}, 014032-1 (2000). 
\bibitem{belle}
Belle collaboration, K.~Abe {\it et al.}, Phys. Rev. D {\bf 69}, 112002 (2004). 
\bibitem{argus}
ARGUS collaboration, H.~Albrecht {\it et al.}, Zeit. Phys. {\bf C~57}, 533 (1993).
\bibitem{cleo}
CLEO collaboration, A.~Anastassov {\it et al.}, Phys. Rev. Lett. {\bf 80}, 4127 (1998). 
\bibitem{opal}
OPAL collaboration, G.~Abbiendi {\it et al.}, Eur. Phys. J. {\bf C~30}, 467 (2003).
\bibitem{aleph}
ALEPH collaboration, D.~Buskolic {\it et al.}, Zeit. Phys. {\bf C~73}, 601 (1997).
\bibitem{delphi} 
DELPHI collaboration, D.~Bloch {\it et al.}, at the XXXth International Conference on 
High Energy Physics, Osaka, DELPHI Report No. 2000-106-CONF (2000).
\bibitem{run2det} 
V.~Abazov {\it et al.}, ``The Upgraded D\O\ Detector'',
in preparation for submission to 
Nucl. Instrum. Methods Phys. Res. A.
\bibitem{eta}
The pseudorapidity $\eta$ is defined using the polar angle with respect to the proton beam direction, 
$\theta$, as $\eta=-\ln[\tan( \theta/2)]$. 
\bibitem{run1det} 
V. Abazov {\it et al.}, ``The Muon System of the Run II D\O\ Detector", 
physics/0503151;
S.~Abachi {\it et al.}, Nucl. Instrum. Methods Phys. Res. A {\bf 338}, 185 (1994). 
\bibitem{footnote}
$B$ refers to the $B^0$ and $B^+$ mesons, and 
charge conjugate states are always implied throughout the Letter.
\bibitem{dt}
D\O\ collaboration, V.~Abazov {\it et al.}, 
Phys. Rev. Lett. {\bf 94}, 182001 (2005).
\bibitem{axial} Axial plane is defined as a plane transverse to the beam direction.
\bibitem{pdg}
Particle Data Group, S.~Eidelman {\it et al.}, Phys.\ Lett. B~{\bf 592}, 1 (2004).
\bibitem{blwe}
J.~Blatt and V.~Weisskopf, Theoretical Nuclear Physics,  John Wiley \& Sons, New York, 1952, p.~361.
\bibitem{evtgen}
D.~J.~Lange, Nucl. Instrum. Methods Phys. Res. A {\bf 462}, 152 (2001).
\bibitem{PYTHIA}
T.~Sj\"ostrand {\it et al.}, Comp. Phys. Commun. {\bf 135}, 238 (2001).
\bibitem{geant}
R.~Brun {\it et al.}, CERN Report No. DD/EE/84-1, 1984.
\bibitem{manohar}
A.~V.~Manohar and M.~B.~Wise, Heavy Quark Physics, 
Cambridge Monogr. Part. Phys. Nucl. Phys. Cosmol. 10 (2000).
\end{thebibliography}
\end{document}